\documentclass[
reprint,
superscriptaddress,
pre
]{revtex4-1}

\usepackage{color}
\usepackage{mathtools}
\usepackage{hyperref}
\usepackage{amsmath,amssymb,amsfonts}
\usepackage{graphicx}
\usepackage{dcolumn}
\usepackage{bm}
\usepackage{float}
\usepackage{epstopdf}
\usepackage{textcomp}
\usepackage{verbatim}

\def\be{\begin{equation}}
\def\ee{\end{equation}}
\def\bea{\begin{eqnarray}}
\def\eea{\end{eqnarray}}
\def\bsplit{\begin{split}}
	\def\esplit{\end{split}}

\def\p{\partial} 

\def\f{\frac}

\def\l{\left(}
\def\r{\right)}

\def\la{\langle}
\def\ra{\rangle}

\def\mr{\mathrm}
\def\refn{Eq.\,\ref}

\newcommand{\curie}{\affiliation{Laboratoire Physico Chimie Curie, Institut Curie, PSL Research University, CNRS UMR168, 75005 Paris, France}}

\begin{document}

	\title{Optimizing Energetic Cost of Uncertainty in a Driven System With and Without Feedback}

	\author{Amit Singh Vishen}
	\thanks{amitsingvishen@gmail.com}
	\curie
	
	\date{October 27, 2020}
	
	\begin{abstract}
		
		Many biological functions require the dynamics to be necessarily driven out-of-equilibrium. In contrast, in various contexts, a nonequilibrium dynamics at fast timescales can be described by an effective equilibrium dynamics at a slower timescale. 
		In this work we study the two different aspects, (i) the energy-efficiency tradeoff for a specific nonequilibrium linear dynamics of two variables with feedback, and (ii) the cost of effective parameters in a coarse-grained theory as given by the ``hidden" dissipation and entropy production rate in the effective equilibrium limit of the dynamics. 
		To meaningfully discuss the tradeoff between energy consumption and the efficiency of the desired function, a one-to-one mapping between function(s) and energy input is required. The function considered in this work is the variance of one of the variables.  We get a one-to-one mapping by considering the minimum variance obtained for a fixed entropy production rate and vice-versa. 
		We find that this minimum achievable variance is a monotonically decreasing function of the given entropy production rate. When there is a timescale separation, in the effective equilibrium limit, the cost of the effective potential and temperature is the associated ``hidden" entropy production rate.
		
	\end{abstract}
	
	\maketitle
	
	\section{Introduction}
	
	Adaptation, kinetic proofreading, and motor transport are a small set of examples of cellular processes that are necessarily nonequilibrium. These processes require a finite rate of energy input to perform the desired function \cite{Gnesotto2017, Seifert2012, Mehta2016}. A natural question is whether an increase in the energy input leads to an increase in the efficiency of the desired function(s) of such nonequilibrium systems?
	
	At the core of a sensory adaptation dynamics is a negative feedback control circuit, for which there is a tradeoff between adaptation error, energy input, and adaptation speed \cite{Lan2012, Sartori2014, Sartori2015, Wang2015}. Similar tradeoffs in other negative feedback circuits exist \cite{Lan2013, De2013}.
	In kinetic proofreading, there is a tradeoff between discrimination efficiency, time, and energy consumption \cite{Ehrenberg1980, Murugan2012}. For the directed motion of motor proteins, the speed depends on the Adenosine triphosphate (ATP) input \cite{Julicher1997, Brown2019}.
	The cell's efficiency in estimating the concentration of ligands improves with an increase in the rate of energy consumption \cite{Mehta2016, Mehta2012, Lang2014, Govern2014, Barato2013}. 
	These studies seem to indicate that the performance of energy-consuming tasks improves on increasing the rate of energy input into the system. 
	However, a recent study shows that, for some of the processes mentioned above,  an increase in the rate of energy dissipation leads to a decrease in the efficiency of the desired function, for some value of parameters \cite{Baiesi2018}. 
	
	Through an example of a linear feedback dynamics, we show that these seemingly contradictory observations are due to lack of a one-to-one mapping between the function(s) and the rate of energy consumption. A meaningful discussion of the energy-efficiency tradeoff requires a well-defined optimization problem, which leads to a one-to-one mapping between the rate of energy consumption and the desired function. As is shown later, this is achieved by first optimizing the function for given energy input and then studying this optimum as the value of the energy input is changed. In an alternative approach, for a class of functions, the ambiguity in energy-efficiency tradeoff can be resolved by the celebrated thermodynamic uncertainty relations (TUR) \cite{Barato2015, Seifert2018, Horowitz2019a}. For example, using TUR, universal bound on the molecular motor efficiency has been obtained \cite{Pietzonka2016}. However, the class of functions to which TUR applies does not cover many of the biologically relevant functions; variance of molecule number, adaptation error, to name a few.

	In contrast to the processes mentioned above, which are necessarily described by a nonequilibrium model, are the processes which, at timescales of interest, can be successfully described by an effective equilibrium model, even though the underlying processes at faster timescales are nonequilibrium. 
	For instance, frequently, the cell cortex is effectively described by effective surface energy, although the underlying dynamics are nonequilibrium due to ATP consumption by actin and molecular motors. The effective tension description has been extensively used to model tissue shapes using vertex models \cite{Fletcher2014, Alt2017}.   
	
	An effective equilibrium limit of a nonequilibrium process has been a subject of various recent studies in active matter \cite{Szamel2014, Fodor2016, Fily2017, Wittmann2017, Wittmann2017a, Wittmann2018}. For a driven dynamic with a time scale separation, the effective equilibrium limit is obtained by integrating the fast degree of freedom. 
	The remaining slow variables describe the resulting effective equilibrium theory with effective parameters like effective temperature and effective potential that retain some memory of the integrated out fast variables. The steady-state entropy production rate (EPR) and heat dissipation rate (HDR) is zero; however, the limiting value of the EPR obtained from the full dynamics, in general, may be nonzero. This difference in EPR and HDR has been referred to as ``hidden" EPR (HEPR) and ``hidden" HDR (HHDR) \cite{Wang2016, Chun2015, Bo2014, Puglisi2010, Esposito2012, Horowitz2017, Shankar2018}. In this work, we interpret the HHDR as the measure of energetic cost for generating the effective parameters. 
	
	In this work, we study two different aspects, (i) the energy-efficiency tradeoff for a specific nonequilibrium model, and (ii) the cost of effective parameters of the coarse-grained theory given by the ``hidden" EPR.
	We analyze the stochastic dynamics of two coupled variables $x$ and $y$ given by Overdamped Langevin equations for two different couplings - feedforward and with feedback. 	
	For both feedforward and feedback dynamics, we analyze the effective equilibrium limit when there is a clear timescale separation in the dynamics, i.e., the dynamics of one of the variables ($y$) is much faster than that of the other ($x$).  
	We consider two models for the separation of timescales, referred to as Model 1 and Model 2 in Ref. \cite{Nandi2018}. 
	For concreteness, the variables $x$ and $y$ are taken to be positions of two point particles. However, the analysis and the results of this work are valid more generally. For instance, the variables $x$ and $y$ may represent concentrations of chemical components, with temperate baths replaced by chemical baths. This is frequently the case for biological signaling networks that are driven by the chemical potential difference of ATP and ADP \cite {Qian2005, Ge2012, Hernandez2012}. 
	
	For the energy-efficiency tradeoff, the function considered in this work is the steady-state variance of variable $x$. 
		In different biological contexts, both increase and decrease of this variance can be the desired goal. 
		For instance, feedforward dynamics lead to an increase in effective temperature, which leads to a faster reaction rate in a diffusion-limited reaction by increasing the diffusion. 
		In contrast, dynamics with negative feedback leads to a decrease in effective temperature; hence, it reduces the variability that can be the desired for protein synthesis \cite{Thattai2001, Paulsson2004, Sneppen2010, Hinczewski2016}. 
		Thus we see that for the same function - variance of $x$, the efficiency depends on the underlying circuit. 
		This is very different than the function - variance of probability current fluctuations for which universal bounds exist, given by the thermodynamics uncertainty relations, that is independent of any specific model \cite{Barato2015, Seifert2018, Horowitz2019a}.

	Following are the main results: (i) the two commonly used models of stochastic forcing in the feedforward process (Model 1 and Model 2 in Ref. \cite{Nandi2018}) lead to very different values of variance and HEPR in white noise limit (see \refn{eq:ff_limit_temp} and \refn{eq:ff_limit_epr}). 
		This result shows the importance of including the correct physical model to compute the HEPR.
		(ii) A one-to-one mapping between the function (variance of $x$) and the cost (HDR and EPR) exists for feedforward dynamics. We find that the variance or effective temperature increases with an increase in EPR at a given timescale of dynamics.  (iii) For dynamics with feedback, a one-to-one mapping between variance and EPR is obtained by minimizing the variance for a given EPR and vice-versa. We find that the minimum variance is a monotonically decreasing function of the given EPR. Equivalently, for a given variance, the minimum EPR required is a monotonically increasing function (see  \refn{eq:minepr2} and \refn{eq:fb_TUR2}). This shows that,  for the model studied, there is a clear tradeoff between the variance and the EPR.
	
	In the following, we compute the correlation between the EPR and the HDR with the variance of $x$, first for the feedforward dynamics in Sec.\ref{sec:ff}, and then for the feedback dynamics in Sec. \ref{sec:fb}.
	
	\section{Particle driven by an Ornstein-Uhlenbeck process}\label{sec:ff}
	
	We first consider a feedforward dynamics where the particle at $x$ is driven by the particle at $y$, and $y$ is an Ornstein-Uhlenbeck process (OUP).  
	The dynamics is taken to be linear, valid for small perturbations around steady-state of a nonlinear dynamics. In this limit the Overdamped Langevin dynamic of particles positioned at $x$ and $y$ is taken to be
	\bea
	\label{eq:langevin_x}
	\dot x  &=&  - \mu_x \l x - \alpha y \r + \sqrt{2 T_x \mu_x} \xi_x(t),\\
	\dot y &=&  - \f{\mu_y}{\tau} y + \f{1}{\tau^{\epsilon}}\sqrt{2 T_y \mu_y}  \xi_y(t), 
	\label{eq:langevin_y}
	\eea
	where $\mu_x$ and $\mu_y$ are the mobilities of particle $x$ and $y$ respectively, $T_x$ and $T_y$ are   temperatures corresponding to the heat bath of particle $x$ and $y$ respectively, and  $\xi_x(t)$ and $\xi_y(t)$ are Gaussian white noise of zero mean and correlation $\la \xi_i(t)\xi_j(t')\ra = \delta_{ij}\delta(t-t')$, where $i,j\in \{x,y\}$. The separation of timescale between $x$ and $y$ is made explicitly by $\tau$. 
	In general, the correlation time of $y$ and its variance may scale differently with $\tau$,  this difference in scaling is captured by $\epsilon$; the mobility scales as $\tau^{-1}$ and the fluctuation scales as $\tau^{1-2\epsilon}$.
	From \refn{eq:langevin_y} the correlation function when $(t+t')\gg \tau $ is $\la y(t)y(t')\ra = \tau^{1-2\epsilon}T_y\exp(-\mu_y|t-t'|/\tau)$. 
	
	The two choices of $\epsilon$ commonly used in the literature are $\epsilon =1/2$ and $\epsilon = 1$, referred to as Model 1 and Model 2, respectively in Ref. \cite{Nandi2018}.
	In Ref. \cite{Wang2016} where the stochastic driving is due to a fluctuating harmonic potential the dynamics under fast switching corresponds to the Langevin \refn{eq:langevin_x} and \refn{eq:langevin_y} with $\epsilon = 1/2$. This is also the case when $y$ is connected to a given temperature bath and the timescale separation is due to the scaling of the mobility $\mu_y$. For $\epsilon = 1/2$  $\lim_{\tau \to 0}\la y(t)y(t')\ra =0$.
	In contrast, most effective equilibrium models of active particles driven by an OUP use $\epsilon =1$ \cite{Fodor2016}. Effective equilibrium theories have been constructed in the limit $\tau\to0$ using the value $\epsilon = 1$, in this case $\lim_{\tau \to 0}\la y(t)y(t') \ra = 2T_y \delta(t-t')$. 
	See Ref.\, \cite{Nandi2018} for the discussion on the implication of difference in choices of $\epsilon$ in the context of glass dynamics.
	
	Using the framework of Stochastic thermodynamics \cite{Seifert2012,Sekimoto2010}, the external work done on the particle $x$ by particle $y$ is given by $\dot w_\mr{ext} = \la \alpha\, y\dot x\ra$. This work is dissipate as heat in the temperature bath $T_x$. The heat flow into the bath $T_x$ can be calculated using one of the standard methods from Stochastic thermodynamics framework \cite{Seifert2012,Harada2005,Harada2006}. In this work we use the Harada-Sasa relation \cite{Harada2005,Harada2006} to compute the HDR and EPR.
	
	The correlation function of $x$ is defined as
	\be 
	\label{eq:corr}
	C_{xx}(t) = \la x(t) - \la x\ra \ra \la x(0) - \la x\ra\ra,
	\ee 
	where (and in the rest of the paper) $\la \cdot \ra$ denotes the ensemble average at steady state.
	From \refn{eq:langevin_x} we get $\la x(t)\ra = 0$. Using Parseval's theorem  \cite{Arfken1999} , the steady state variance of $x$ is given by
	\be
	\label{eq:var1}
	\la x^2\ra_\mr{ff} =  \int_{-\infty}^{\infty} \f{d\omega}{2\pi}\tilde C_{xx}(\omega),
	\ee  
	where the tilde denotes the Fourier transform defined as $\tilde \phi(\omega) = \int_{-\infty}^{\infty} dt\, e^{-i\omega t}\phi(t)$. The subscript ``$\mr{ff}$" denoting feedforward dynamics.
	Obtaining the correlation spectrum  from \refn{eq:langevin_x}, \refn{eq:langevin_y} and using \refn{eq:var1} we get the steady state variance of $x$ as
	\be
	\label{eq:ff_var1}
	\la x^2\ra_\mr{ff} = T_x + \f{\mu_x \alpha^2 T_y \tau^{2 - 2 \epsilon}}{\mu_y + \mu_x \tau},
	\ee
	where the first term on the right is the direct contribution due to temperature bath $T_x$ and the second term is due to the coupling to temperature bath $T_y$.  
	
	The linear response function for a small force $\delta f_p$  is defined as \cite{Chaikin2000}
	\be 
	\label{e:resp}
	\la x \ra_{\delta} - \la x\ra = \int \chi_x(t-t')\delta f_p(t') dt',
	\ee 
	where $\la \cdot \ra_{\delta}$ denotes the ensemble average over the steady-state of the perturbed dynamics. Using \refn{eq:langevin_x} and \refn{eq:langevin_y} we get the response function in frequency space as $\tilde \chi_x =  1/(-i\omega+\mu_x)$.

	From the response and correlation function the heat dissipation rate can be obtained using the Harada-Sasa, which, for zero mean velocity reads as \cite{Harada2005,Harada2006}
	\be
	\label{eq:hdr1}
	h_x = \f{1}{\mu_x}   \int_{- \infty}^{\infty} \f{d\omega}{2\pi} \l \omega^2 \tilde C_{xx}(\omega) -  2 \omega T_x \tilde \chi''_{x}(\omega)\r,
	\ee
	where we have used the relation $\tilde C_{vv} = \omega^2 \tilde C_{xx}$ and $\tilde \chi'_{v} =  \omega \tilde\chi_{x}''$. Note that we have used $T_x$ as the temperature of the heat bath connected to $x$. In case $T_x$ includes athermal contributions then the corresponding fast degrees-of-freedom need to be included to obtain correct EPR and HDR.  Since there is no feedback of $x$ on $y$ the HDR corresponding to variable $y$ is zero. The heat flow into bath $T_x$ is the work done by the the variable $y$ on $x$. Note that, this work-done is due to an external energy source, apparent from the non-conservative coupling between $x$ and $y$ .
	The EPR is given by $\sigma = h_x/T_x$. Using \refn{eq:hdr1} we get 
	\be 
	\label{eq:epr1}
	\sigma = \f{T_y \mu_y \mu_x \alpha^2 \tau^{1 - 2\epsilon  }}{T_x (\mu_y + \mu_x \tau)}.
	\ee 
	Without the form of explicit timescale separation $\tau$ this EPR has been obtained in various other contexts \cite{Wang2016,Shankar2018,PrawarDadhichi2018,Chaki2019}. 
	
	\noindent{\bf Fluctuation and dissipation rate}: From \refn{eq:ff_var1} and \refn{eq:epr1} we get a one-to-one mapping between the variance of $x$ and entropy produced over the timescale $\tau/\mu_y$,
	\be 
	\f{\la x^2\ra_\mr{ff}}{T_x} = 1 + \f{\tau}{\mu_y}\sigma.
	\ee 
	Thus we see that, for the feedforward dynamics, increasing EPR at a given timescale $\tau/\mu_y$ leads to an increase in the variance of $x$. If the goal is an increase of variance then it does seem to hold an energy-efficiency trade-off, more energy needs to be spent to attain higher variance. 
	However, if the goal is the reduction of variance, then it seems that spending more energy leads to less efficiency. Indeed, the minimum variance and EPR is for $\alpha = 0$.
	This highlights again the importance of making a correct guess for the function corresponding to the given dynamics.  
	
	\noindent{\bf Effective equilibrium limit}: In the limit $\tau \to 0$ and $\epsilon \leq 1$ (for $\epsilon > 1$ the adiabatic limit does not exist) the two-variable nonequilibrium dynamics reduces to  the following effective equilibrium dynamics: 
	\be
	\label{eq:eff_x}
	\dot x = - \mu_x x + \sqrt{2T_\mr{eff}\mu_x}\, \xi_x,
	\ee 
	where the effective temperature is given by
	\bea 
	\label{eq:ff_limit_temp}
	\f{T_\mr{eff}}{T_x} = \left\{ \begin{array}{c c}
		1 & \epsilon < 1,\\
		1 + \f{\mu_x \alpha^2 T_y}{\mu_yT_x} & \epsilon =1.
	\end{array} \right.
	\eea 
	The variance $\la x^2\ra_\mr{ff} = T_\mr{eff}$.
	The EPR corresponding to \refn{eq:eff_x} is zero; however, $\Delta \sigma = \lim_{\tau \to 0} \sigma \neq 0$, where $\Delta \sigma$ is the HEPR given by
	\bea 
	\label{eq:ff_limit_epr}
	\f{\Delta \sigma}{\mu_x} = \left\{ \begin{array}{c c}
		0 & \epsilon < 1/2,\\
		T_y \alpha^2 /T_x & \epsilon = 1/2, \\
		\infty & \epsilon > 1/2.
	\end{array} \right.
	\eea 
	We see that $\Delta \sigma$ is finite only for $\epsilon \leq 1/2$.  
	Consistent with Ref. \cite{Wang2016} we get finite HEPR for $\epsilon = 1/2$, which depends on the parameters of the fast variable even though  effective dynamics in  \refn{eq:eff_x} does not.  
	If $\epsilon > 1/2$,  we need to include faster degrees of freedom like inertial relaxation; this introduces a high-frequency cut-off leading to a finite HEPR \cite{Shankar2018}.
	For $\epsilon  = 1/2$ the effective temperature to leading order in $\tau$ reads
		\be 
		\f{T_\mr{eff}}{T_x} =  1 + \tau \f{\mu_x \alpha^2 T_y}{\mu_yT_x},
		\ee 
		the second term on the right is the increase in effective temperature due to the nonequilibrium driving, and the HEPR to leading order in $\tau$ reads
		\be 
		\f{\Delta \sigma}{\mu_x} = \f{T_y \alpha^2}{T_x} \l1 + \tau \f{\mu_x}{\mu_y} \r,
		\ee 
		where the second term on the right can be seen as the additional energetic cost for increasing the effective temperature.

	\section{With feedback}\label{sec:fb}
	
\begin{figure}
		\centering
		\includegraphics[width=\linewidth]{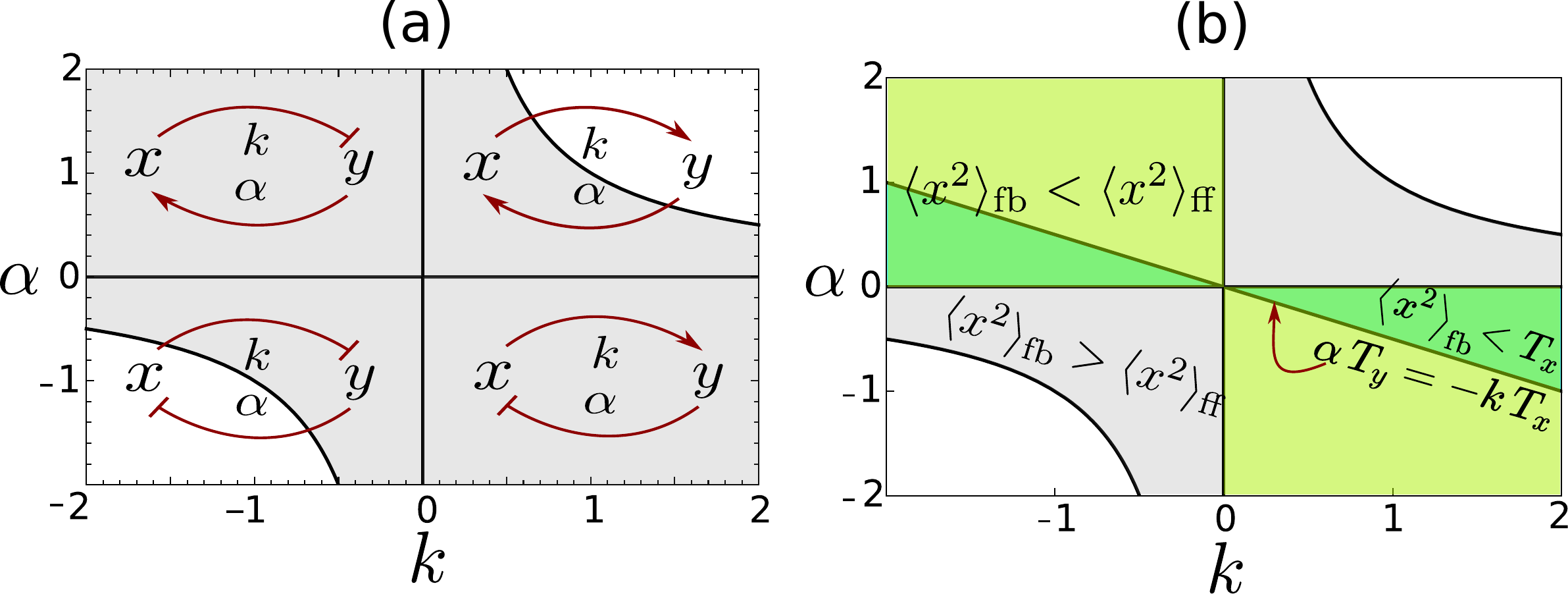}
		\caption{(a) Plot of the stable region in the $k$ and $\alpha$ parameter space, the dynamics is stable for $k\,\alpha < 1$ (shown in gray), and the schematic of the feedback circuit in different quadrants. The feedback is positive for $k\,\alpha>0$ and negative for $k\,\alpha<0$. (b) Plot of the variance of $x$ as function of $k$ and $\alpha$.  When the feedback is positive, $\la x^2\ra_\mr{fb} > \la x^2\ra_\mr{ff}$, and when the feedback is negative,$\la x^2\ra_\mr{fb} < \la x^2\ra_\mr{ff}$.  In the region between the line $\alpha =0$ and $\alpha T_y + kT_x = 0$ the variance of $x$ is less that that without its coupling with $y$, i.e., $\la x^2\ra_\mr{fb} < T_x$). 
		}
		\label{fig:fb1}
	\end{figure}

	We now consider the dynamics with feedback. 
	As before, the dynamics is linearized around the steady state of the non-linear dynamics.  The  stochastic dynamics reads
	\bea
	\label{eq:langevin_x2}
	\dot x &=&  - \mu_x \l x - \alpha y \r + \sqrt{2T_x \mu_x} \xi_x(t), \\
	\tau \dot y &=&  - \mu_y \l  y - k x \r + \tau^{1-\epsilon}\sqrt{2T_y \mu_y} \xi_y(t),
	\label{eq:langevin_y2}
	\eea
	where, as before, $\xi_x(t)$ and $\xi_y(t)$ are Gaussian white noise of zero mean and unit variance, $\alpha$ and $k$ are the feedback parameters, and the separation of timescale is made explicit through $\tau$ and $\epsilon$. 
	The dynamics is stable for $k\,\alpha < 1$.
	Fig.\,\ref{fig:fb1}(a) shows the stable regions and the schematic of the feedback in $k$ and $\alpha$ parameter space. 
	For $k\,\alpha>0$ the feedback is positive and for $k\,\alpha<0$ the feedback is negative. 
	
	Of particular interest is the case of negative feedback that is relevant for various biological function, like noise reduction in protein synthesis \cite{Thattai2001, Paulsson2004, Hinczewski2016}, homeostasis \cite{Sneppen2010}, and adaptation \cite{Lan2012}.
		Since, velocity is odd under time reversal where as the variable $x$ and $y$ are even under time-reversal, the stochastic thermodynamic analysis of \refn{eq:langevin_x2} and \refn{eq:langevin_y2} differs form that of velocity dependent negative feedback studies of molecular refrigeration \cite{Kim2004,Kim2007,Mnakata2012,Munakata2013} and of particle dynamics in a viscoleastic medium \cite{Vishen2020}.
	
	The steady state variance of $x$ after substituting correlation spectrum obtained from \refn{eq:var1} and using \refn{eq:langevin_x2}-\ref{eq:langevin_y2} is
	\be 
	\label{eq:fb_var_x}
	\la x^2\ra_\mr{fb} = \f{T_x\mu_x\tau}{\mu_y + \mu_x \tau} + \f{ \mu_yT_x + \alpha^2\tau^{2-2\epsilon}T_y\mu_x}{(1-k\alpha)  (\mu_y + \mu_x \tau)}.
	\ee 
	Fig.\,\ref{fig:fb1}(b) shows the variance of x as a function of $k$ and $\alpha$ for $\tau =1$. The various regions of the parameter space are discussed in the Sec. \ref{ss:FDR}.
	The HDR corresponding to variable $x$ is given by \refn{eq:hdr1} which upon substituting correlation and response functions obtained from \refn{eq:langevin_x2} and \refn{eq:langevin_y2} gives
	\be 
	\label{eq:fb_heat_x}
	h_{x}^\mr{fb}  = \f{\mu_y\mu_x}{\mu_y + \mu_x \tau}\l \tau^{1-2\epsilon}T_y\alpha^2 - k\, \alpha\, T_x\r.
	\ee 
	The first term in the bracket is the HDR into the bath $T_x$ due to the driving by the bath $T_y$, and the second term is the HDR due to feedback. 
	
	In contrast to feedforward case, the HDR corresponding to the variable $y$ is now nonzero. Taking the effective temperature corresponding to $y$ as $\tau^{1-2\epsilon}T_y$, the HDR of $y$ using Harada-Sasa relation is given by
	\be 
	\label{eq:hdr_y}
	h_y = \f{\tau}{\mu_y}   \int_{- \infty}^{\infty} \f{d\omega}{2\pi} \l \omega^2\tilde C_{yy}(\omega) -  2 \omega T_y\tau^{1-2\epsilon} \tilde \chi''_{y}(\omega)\r.
	\ee
	Substituting the correlation and the response function obtained from \refn{eq:langevin_x2}-\ref{eq:langevin_y2} into \refn{eq:hdr_y} after integration gives
	\be 
	\label{eq:fb_heat_y}
	h_{y}^\mr{fb}  = \f{\mu_y\mu_x}{\mu_y + \mu_x \tau}\l T_x k^2 - k \alpha T_y\tau^{1-2\epsilon}\r.
	\ee 
	Similar to $h_x^\mr{fb}$, the first term in the bracket is the dissipation rate due to the driving by the bath $T_x$ and the second term is due to the feedback. 
	The total HDR $h^\mr{fb} = h_x^\mr{fb} + h_y^\mr{fb}$,
	upon substituting \refn{eq:fb_heat_x} and \refn{eq:fb_heat_y} is given by
	\be 
	\label{eq:fb_heat}
	h^\mr{fb} = \l 1- \f{k}{\alpha}\r h_x^\mr{fb}, 
	\ee 
	and the total EPR is given by
	\be 
	\sigma^\mr{fb} = \f{h_x^\mr{fb}}{T_x} +\f{h_y^\mr{fb}}{T_y\tau^{1-2\epsilon}},
	\ee 
	which upon substituting \refn{eq:fb_heat_x} and \refn{eq:fb_heat_y} gives
	\be 
	\label{eq:fb_epr}
	\sigma^\mr{fb} = \f{\tau^{2\epsilon-1}\mu_y \mu_x (T_x k - \tau^{1-2\epsilon}T_y \alpha)^2}{T_xT_y(\mu_y + \mu_x\tau)}.
	\ee  
	
	\subsection{Conservative vs. Non-Conservative Coupling}
	
	The dynamics given by \refn{eq:langevin_x2} and \refn{eq:langevin_y2} are nonequilibrium due to the non-conservative coupling between $x$ and $y$ and the difference in temperature of the two baths ($T_x \neq T_y$).
	The dynamics for conservative coupling is of the form 
	\bea
	\dot x &=&  \mu_x \p_x \Phi(x,y) + \sqrt{T_x \mu_x} \xi_x(t), \\
	\dot y &=&  \f{\mu_y}{\tau} \p_y \Phi(x,y) + \f{1}{\tau^{\epsilon}}\sqrt{T_y \mu_y} \xi_y(t).
	\eea
	For $T_x = T_y$, the steady state is given by the Boltzmann distribution $P(x,y)\propto e^{-\beta \Phi(x,y)}$ and the HDR, EPR are zero. For $T_x \neq T_y$ there is a finite EPR and HDR at steady state; however, the total heat flow must be zero because the heat flows from the ``hotter" to the ``colder" bath \cite{Fogedby2011,Visco2006}. 
	In \refn{eq:langevin_x2} and \refn{eq:langevin_y2} the coupling is conservative only when $k = \alpha$ for which the potential is $\Phi = x^2/2 + y^2/2 - k\,x\,y $. In this case, from \refn{eq:fb_heat} we get $h^\mr{fb} = 0 $.
	When the coupling is non-conservative, i.e., $k\neq\alpha$, the total HDR $h^\mr{fb} \neq 0 $.
	The nonzero heat flow implies that external variables driving the system are implicit in the dynamics through the non-conservative coupling. This external driving acts as a work reservoir; for $h^\mr{fb} > 0$, there is a net work done on the system  and for $h^\mr{fb} < 0$, the work is being extracted from the system.
	
\begin{figure}
		\centering
		\includegraphics[width=0.7\linewidth]{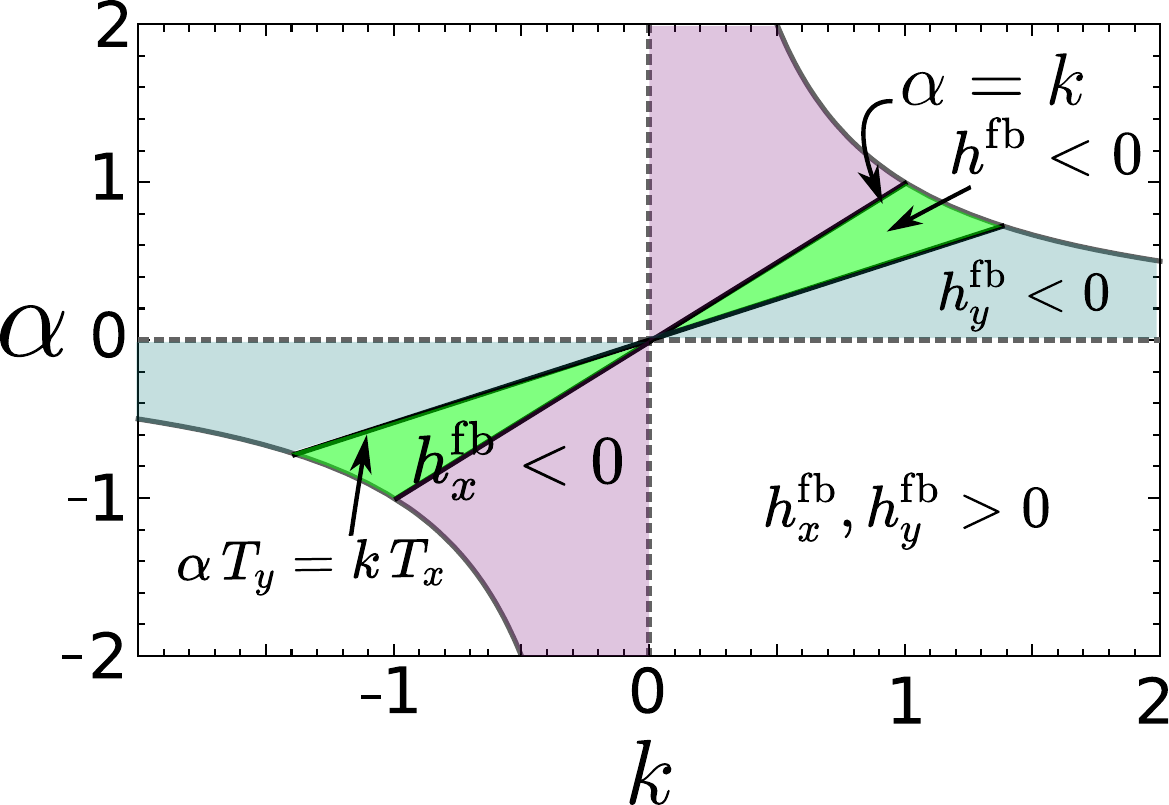}
		\caption{The plot shows the sign of the heat flow into the temperature baths as function of the feedback parameters. When the feedback is negative the heat dissipation rate for heat baths associated with $x$ and $y$ is positive. For positive feedback, in the region between the lines $\alpha T_y = k T_x$ and $\alpha = 0$ ($k=0$) $h_y^\mr{fb}<0$ ($h_x^\mr{fb}<0$), and in the region between the lines $k=\alpha$ and $\alpha T_y = kT_x$ the total heat flow is negative($h^\mr{fb}<0$).
		}
		\label{fig:fb_heat}
	\end{figure}

	Fig.\,\ref{fig:fb_heat} shows the sign of total heat flow. 
	The total heat $h^\mr{fb} =0$ when the coupling is conservative ($k=\alpha$) and when the dynamics can be mapped to an effective equilibrium dynamics ($T_y \alpha = T_x k$). In the region between these two lines, $h<0$ (shown in green), i.e., there is a net heat flow out of the system. In the rest of the parameter space, there is net heat flow into the system. 
	As we can see from \refn{eq:fb_epr}, independent of the sign of $h_x^\mr{fb}$, $h_y^\mr{fb}$, and $h^\mr{fb}$, the EPR given by \refn{eq:fb_epr} is always positive. However, if only one of the variables is considered, say $x$, then the corresponding EPR as given by $h_x^\mr{fb}/T_x$ will be negative when the HDR is negative \cite{Bhattacharya2017, Strasberg2019}.

	Some physical examples of non-conservative coupling are hydrodynamic interaction in the presence of nonequilibrium fluctuations \cite{Berut2014, Berut2016, Berut2016a, Vishen2019}, effective interaction between chemically interacting particles \cite{Agudo2019}, signaling networks, and gene networks. For more discussion, see Ref. \cite{Loos2019}.
	
	\subsection{Effective equilibrium limit}
	Similar to feedforward case, in the limit $\tau\to 0$ and $\epsilon \leq 1$ we get the following effective one-variable dynamics:
	\be 
	\label{eq:fb_eff_x}
	\dot x = -\mu_x(1-k\,\alpha)x + \sqrt{T_\mr{eff}\mu_x} \xi_x,
	\ee 
	where $T_\mr{eff}$ is given by \refn{eq:ff_limit_temp}.
	The effect of feedback is apparent in the stiffness of the harmonic potential and the effective temperature. The steady state variance of $x$ as obtained from \refn{eq:fb_eff_x} is
	\be 
	\label{eq:fb_effvar}
	\lim_{\tau \to 0} \la x^2\ra_\mr{fb} = \f{T_\mr{eff}}{(1-k\,\alpha)}.
	\ee 
	The EPR corresponding to \refn{eq:fb_eff_x} is zero; however, the ``hidden" EPR $\Delta \sigma = \lim_{\tau \to 0} \sigma^\mr{fb} \neq 0$. Using \refn{eq:fb_epr} we get
	\bea 
	\label{eq:fb_hepr}
	\f{\Delta \sigma}{\mu_x} = \left\{ \begin{array}{c c}
		\infty & \epsilon \neq 1/2,\\
		\f{ (T_x k - T_y \alpha)^2}{T_xT_y} & \epsilon = 1/2 .
	\end{array} \right.
	\eea 
	In the following, we focus on the physically relevant case of $\epsilon = 1/2$. Unlike the feedforward case, the variance in the effective theory depends upon the parameters of the integrated out variable. 
	We see that for a given timescale $\mu_x$, there is no one-to-one mapping between the EPR and the variance. For a given EPR, the variance can be tuned by changing the feedback parameters.   
	However, a one-to-one mapping is obtained when the question is set as a well-defined optimization problem. 
	What is the minimum value of variance for a given entropy production rate and timescale? 
	
	The minimum value of variance for a given value of HEPR obtained by minimizing \refn{eq:fb_effvar} for $\Delta \sigma$ given by \refn{eq:fb_hepr} (for $\epsilon = 1/2$) is 
	\be 
	\label{eq:minvar1}
	\Lambda^* = \f{4}{4+ S^*},
	\ee 
	where $S^* = \Delta \sigma/\mu_x$, and $\Lambda^* = \min(\la x^2\ra_\mr{fb}/T_x)$. Thus we see that for given effective temperature $T_x$ there is a one-to-one mapping between the minimum of the variance and $\Delta \sigma/\mu_x$. Specifically, $\mr{min}\l\la x^2\ra_\mr{fb}\r$ is a monotonically decreasing function of the HEPR. Inverting \refn{eq:minvar1} we get $S^*= 4\l 1/\Lambda^* -1 \r$, for $\Lambda^* < 1$; this gives the relation between the minimum EPR required to attain a variance $\Lambda^*$.
	For $\Lambda^*\geq1$, the HEPR $S^* = 0$. 
	The HHDR in the medium is obtained by taking the limit  $\tau\to 0$ in \refn{eq:fb_heat}. The minimum HHDR defined as $H^* = \mr{min}(\lim_{\tau\to 0}h^\mr{fb}/\mu_xT_x)$  required to achieve variance $\Lambda^*$ is
	\be 
	\label{eq:minhdr1}
	H^*= \f{(1+\sqrt{T_y/T_x})^2}{4}S^*,
	\ee 
	thus we see that the HHDR depends on the temperature of bath driving the fast variable $T_y$. 
	\refn{eq:minvar1} can be written as the following inequality
		\be 
		\label{eq:fb_TUR1}
		\f{\la x^2\ra_\mr{fb}}{T_x} \geq \f{4}{4+ \Delta \sigma/\mu_x}.
		\ee 
	
	\subsection{Fluctuation and the Dissipation rate}\label{ss:FDR}

	\begin{figure}
		\centering
		\includegraphics[width=\linewidth]{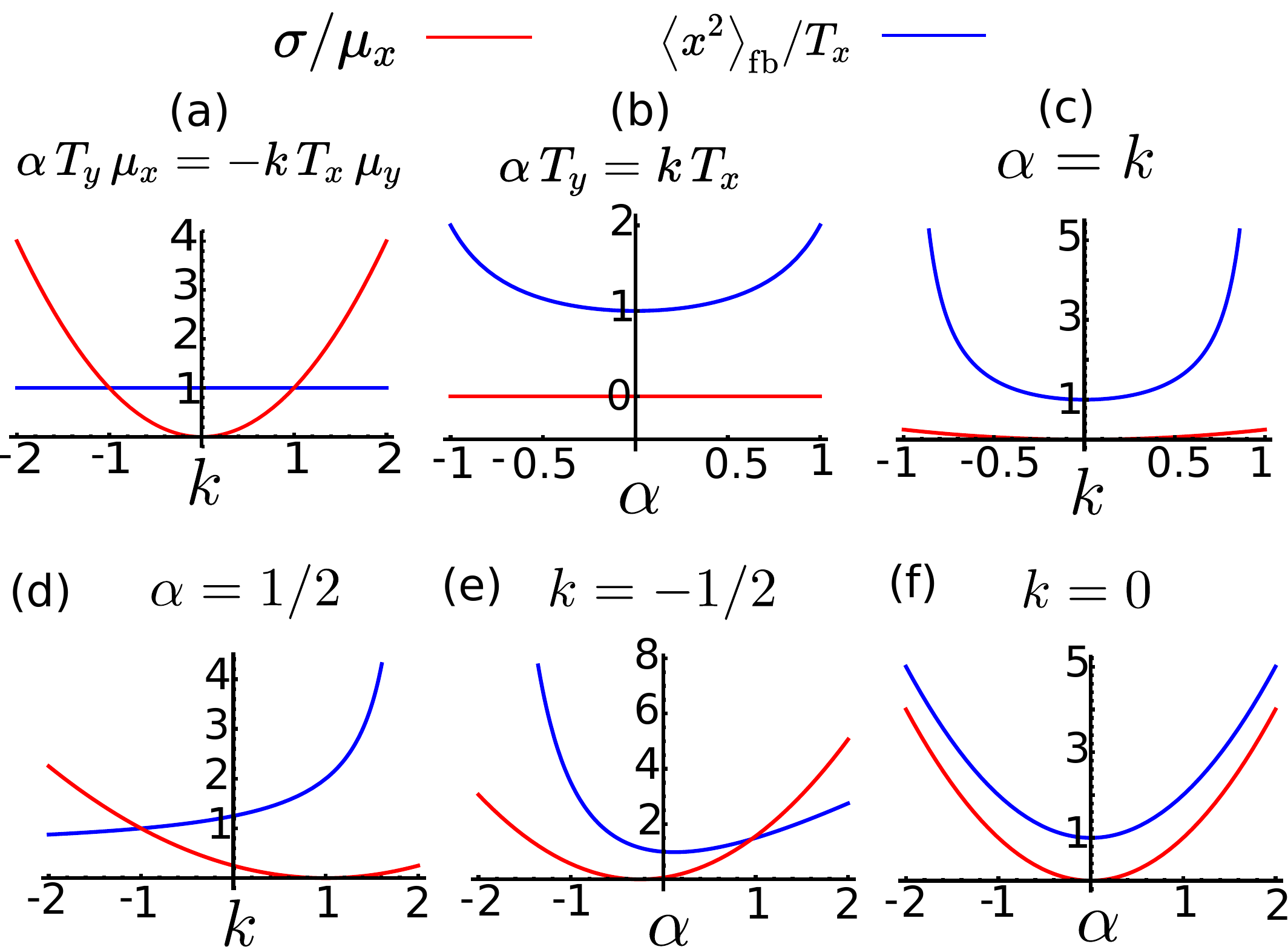}
		\caption{Plot of the entropy production rate and the variance as given by \refn{eq:fb_epr2} and \refn{eq:fb_var_x2} of the main text, for $d=1$, $r=2$, along different section of the parameter space. For (a) $\alpha T_y\mu_x = -kT_x\mu_y$ the variance is constant and equal to that without the coupling with $y$, i.e,$\la x^2\ra_\mr{fb} = T_x$, whereas the EPR depend upon the feedback parameters. In contrast, along (b) $\alpha T_y = k T_x$ the dynamics is at an effective equilibrium ($\sigma = 0$) and the variance depend upon the parameter values. For (c) conservative coupling ($\alpha = k$) and when the coupling is (f) feedforward ($k=0$), both EPR and variance  are correlated and depend upon the parameter values. In general, as show in (d) and (e) the variance and EPR can show diverse trend. 
		}
		\label{fig:fb_sec}
	\end{figure}
	
	To analyze the general case, where there is no timescale separation, we set $\tau = 1$. 
	Using the definition $r \equiv T_y/T_x$ and $d \equiv \mu_x/\mu_y$ \refn{eq:fb_var_x} reads
	\be 
	\label{eq:fb_var_x2}
	\f{\la x^2\ra_\mr{fb}}{T_x} = \f{(1+ d) + d\,\alpha(\alpha\, r-k)}{(1-k\,\alpha)  (1 + d)} .
	\ee 
	It can be shown that the variance $\la x^2\ra_\mr{fb}$ is a monotonically increasing function of $r$ and $k$, and a monotonically decreasing function of $d$. The variance is non-monotonic in $\alpha$, the minimum of $\la x^2\ra_\mr{fb}$ obtained from \refn{eq:fb_var_x2} is at $
	\alpha = 1/k - \sqrt{(k^2 + d\,r)/k^2 d\,r}$.
	Fig.\,\ref{fig:fb1}(b) show the effect of feedback on the variance. 
	As is well established \cite{Thattai2001,Paulsson2004}, negative feedback leads to reduction of variance,  i.e., $\la x^2\ra_\mr{fb} < \la x^2\ra_\mr{ff}$. 
	Moreover, in a sub-region of the negative feedback parameter space, between the line $\alpha T_y \mu_x = -k T_x \mu_y $ and $\alpha = 0$, the variance of $x$ with feedback is less than that without any coupling with $y$, i.e., $\la x^2\ra_\mr{fb}<T_x$. 
	In the stable regions of the first and third quadrant the feedback is positive ($k\,\alpha > 0$) and the fluctuation is larger than that without feedback, i.e., $\la x^2\ra_\mr{fb} > \la x^2\ra_\mr{ff}$. 
	The EPR  in \refn{eq:fb_epr} for $\tau = 1$ reduces to
	\be 
	\label{eq:fb_epr2}
	\f{\sigma^\mr{fb}}{\mu_x} = \f{(k - r \alpha)^2}{r(1 + d)}.
	\ee 
	The minimum value of the EPR is $\sigma^\mr{fb}=0$. The EPR is zero along the line $\alpha T_y = k T_x$, for this value the fluctuation dissipation relation is satisfied; hence, the dynamics can be mapped to the following effective equilibrium model
	\bea
	\label{eq:langevin_x3}
	\dot x &=&  - \mu_x \l x - \alpha y \r + \sqrt{2T_x \mu_x}\, \xi_x(t), \\
	\dot y &=&  - \mu  \l  r y - \alpha x \r + \sqrt{2T_x\mu}\, \xi_y(t),
	\label{eq:langevin_y3}
	\eea
	where $\mu = \mu_yT_y/T_x$. 
	This mapping is not unique, scaling of the mobility and the effective temperature leads to equivalent models with identical steady-state distribution.
	The total heat dissipation rate from \refn{eq:fb_heat} is given by
	\be 
	\label{eq:fb_heat_x2}
	\f{h^\mr{fb}}{\mu_xT_x}  = \f{(\alpha - k)\l r\,\alpha - k\r}{(1+d)}.
	\ee 
	\noindent{\bf Energy-efficiency trade-off}:
	Does increase in EPR lead to a decrease in the variance? 
	Fig.\,\ref{fig:fb_sec} shows the fluctuation and the EPR along different sections of the parameter space.  
	Along the line $\alpha T_x \mu_y = - kT_y \mu_x$, the variance is constant $\la x^2\ra_\mr{fb} = T_x$ but the EPR depends on the feedback parameters, the minimum of EPR is at $k = \alpha = 0$ (Fig.\,\ref{fig:fb_sec}(a)).
	In contrast to this, along the line $T_y \alpha = k T_x$, the EPR is zero but the variance depends on the feedback parameters, the minimum of variance is at $k = \alpha$ (Fig.\,\ref{fig:fb_sec}(b)). 
	In general, we can find regions in the parameter space where the variance decreases with an increase in the EPR as well as regions where the variance increases with an increase in the EPR (Fig.\,\ref{fig:fb_sec}(d,e)). 
	
\begin{figure}
		\centering
		\includegraphics[width=\linewidth]{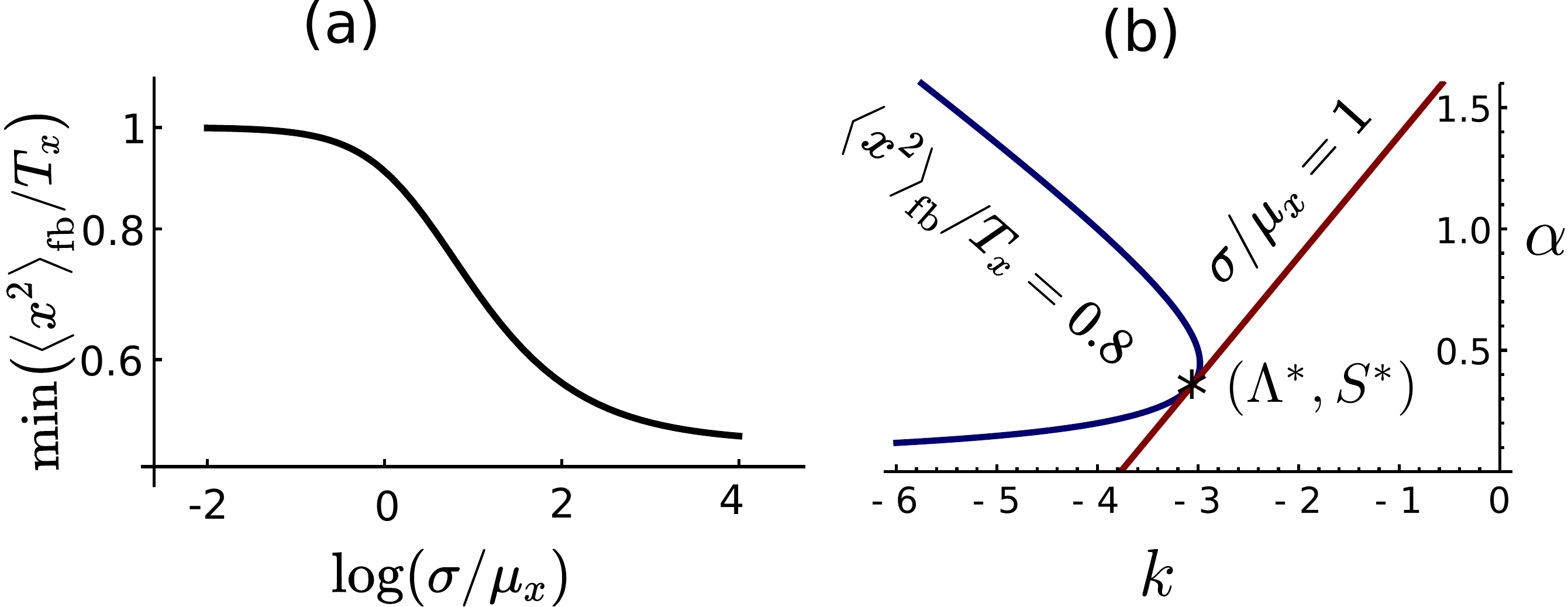}
		\caption{(a) Plot of the minimum value of the variance for a fixed EPR and timescale as function of the given EPR and timescale, for $d=1$ and $r=2$, and (b) the plot showing the curves with constant value of variance (blue) and constant entropy production rate (red), the minimum value of the variance for a given EPR and vice-versa is given by the intersection of the two curves at $\Lambda^*$ and $S^*$.
		}
		\label{fig:epr_var}
	\end{figure}

	As shown in the effective-equilibrium limit, there is a one-to-one mapping between the minimum variance for a given EPR and vice-versa. 
	For $\mr{min}(\la x^2\ra_\mr{fb}) > T_x$ the minimum EPR required is zero. 
	For $\mr{min}(\la x^2\ra_\mr{fb}) < T_x$ a finite minimum EPR is required. The minimum value of EPR is obtained by minimizing the function in \refn{eq:fb_epr2} for a variance  given by \refn{eq:fb_var_x2}. 
	After minimization we get
	\be 
	\label{eq:minepr2}
	S^* = \f{4\Lambda^*(1-\Lambda^*)(1+d)}{\l(1+d)\Lambda^* - d\r^2}
	\ee 
	where $S^* = \sigma/\mu_x$, $\Lambda^* = \mr{min}(\la x^2\ra_\mr{fb})/T_x$. For $d\to0$ this reduces to the HEPR in one-variable limit, given by  \refn{eq:minvar1}.
	The minimum fluctuation for a given EPR is obtained by inverting this equation. This inverted function is plotted in 	Fig.\,\ref{fig:epr_var}(a), we see that the minimum variance is a monotonically decreasing function of the EPR. Thus we see that, in this particular case, an increase in the EPR budget leads to a decrease in the variance.
	The limiting values are
	$\lim_{S^*\to 0} \Lambda^* = T_x \,\, \mr{and} \,\, \lim_{S^*\to \infty} \Lambda^* = dT_x/(1+d)$.
	
	The value of parameters for which the minimum variance is attained are
	\be 
	\label{eq:alpha}
	\alpha^* = \f{\sqrt{(1+d)}\l-1\pm \sqrt{1+d+d\,S^*}\r}{\sqrt{d^2\,r\,S^*}},
	\ee 
	and $k^* = r\alpha^* + \sqrt{(1+d)r\,S^*}$. 
	 The contour of minimum EPR $S^*$ is the line $k = r\alpha \pm \sqrt{S^*r(1+d)}$. In the region between the origin and this line, the EPR is less than $S^*$; hence, the line of  EPR $S^*$ has to be tangent to the contour of given variance $\Lambda^*$ (see Fig.\,\ref{fig:epr_var}(b)). This can be easily verified by calculating the tangent to the curve in \refn{eq:fb_var_x2} for $\Lambda^* = \la x^2\ra_\mr{fb})/T_x$ at point  $k^*$ and $\alpha^*$ as given by \refn{eq:alpha}.
	
	We emphasize that here we have taken the constraint to be the EPR. However, the constraint could very well be the total HDR. For $r =1$ the $HDR$ and $EPR$ are proportional; hence the minimum HDR required to attain a given variance is $H^* = T_x S^*$. When $r\neq1$ the minimum dissipation can be obtained by minimizing \refn{eq:fb_heat_x2} for a variance given by \refn{eq:fb_var_x2}. 
	We can express 
	\refn{eq:minepr2} as an inequality of the from
		\be 
		\label{eq:fb_TUR2}
		\f{\sigma}{\mu_x} \geq \f{4\la x^2\ra_\mr{fb}(T_x-\la x^2\ra_\mr{fb})(1+d)}{T_x\l(1+d)\la x^2\ra_\mr{fb} - d\,T_x\r^2}.
		\ee	
		Note that this inequality is not of the standard form obtained in thermodynamic uncertainty relations (TUR) \cite{Barato2015,Seifert2018,Horowitz2019a}. This is not surprising since the variance is even under time reversal; it is well defined for equilibrium dynamics. For functions like motor efficiency TUR's can be used which provide a fundamental bound, independent of the details of the model \cite{Pietzonka2016}.

	\section{Discussion}
	
	In summary, we consider two interacting particles $x$ and $y$ that are driven out-of-equilibrium by non-conservative forces and connected to different temperature baths. 
	We calculate the steady-state variance of $x$, heat dissipation, and entropy production rate when the coupling between the particles is feedforward, and when there is feedback. 
	
	An effective one-particle description is obtained when there is a separation of timescales between the dynamics of the two particles. 
	In this limit, the parameters of the slow variable $x$ depend upon its coupling with the integrated out fast variable $y$. For the feedforward case, the effective parameter is the effective temperature. 
	In the presence of feedback, the effective theory is described by an effective potential and an effective temperature. 
	The ``hidden" entropy production rate for the two cases with and without feedback depends upon the relative scaling of the temperature $T_y$ and the mobility $\mu_y$.   
	The HEPR is the cost associated with the effective parameters in the coarse-grained equilibrium theory. 
	When the feedback is negative, the large stiffness of the effective potential (smaller variance) requires larger HHDR and HEPR. 
	
	In the absence of timescale separation, the variance of $x$, the EPR, and the HDR depend on the ratio of the mobilities ($d = \mu_x/\mu_y$). The lower bound on the variance is set at $T_x d/(1+d)$. 
	Negative feedback is always nonequilibrium, and for suitable values of the parameter, it leads to a reduction of variance in comparison to the independent dynamics. 
	
	Does an increase in energy dissipation always lead to an improvement in function (variance of $x$)? 
	We find that for a given timescale, the relation between EPR and variance could be very heterogeneous. For instance, the EPR can be changed without affecting the variance and vice-versa. 
	A similar observation has been made in Ref.\,\cite{Baiesi2018}, which contradicts the results in Ref.\,\cite{Lan2012}, the later shows a tradeoff between speed-energy-error and the former shows that the efficiency does not always improve with an increase in energy. 
	In this paper, we argue that the tradeoff problem in these studies is ill-posed since there is no one-to-one mapping possible between function(s) and energy consumption without setting up a well-defined optimization problem. This is even more obvious for a higher dimensional problem involving more variables and parameters. 
	
	A one-to-one mapping between the energy dissipation and efficiency is obtained by minimizing the dissipation as a function of variance or vice-versa. 
	We find that there is a minimum entropy production rate required to decrease the variance below its value in the absence of feedback. This minimum value increases with a decrease in the variance. 
	Thus for the reduction of fluctuation by negative feedback, the more the energy input, the lower the minimum variance. 
	However, it is far from clear that any function which requires the dynamics to be necessarily nonequilibrium leads to such energy-efficiency relation. A more general analysis in a higher dimension should be a useful future direction to explore. 
	The results obtained in this paper should be useful in understanding the evolutionary trajectory of the biological signaling and gene networks that have evolved for improved efficiency under energetic constraints  \cite{Wohlgemuth2011, Ilker2019}.
	
	\section{Acknowledgments}
	ASV thanks J.-F. Joanny and Efe Ilker for helpful discussions, and S.C. Al-Izzi for critical reading of the manuscript. The author is thankful to the referees for their valuable feedback. 
	This work received support from the LabExCelTisPhyBio (ANR-11-LABX-0038).

\end{document}